\documentclass[12pt,twocolumn]{article}

\usepackage[english]{babel}
\usepackage[T1]{fontenc}
\usepackage{cancel}
\usepackage{color}
\usepackage[svgnames]{xcolor}
\usepackage{amssymb,amsmath}
\usepackage{graphicx}
\usepackage{dsfont}
\usepackage{colortbl}
\usepackage{hyperref}
\usepackage[affil-it]{authblk}
\usepackage{subfig}
\usepackage{gensymb}
\usepackage{abstract}
\usepackage[a4paper,total={180mm,260mm}]{geometry}

\AtBeginDocument{}

\hypersetup{
    colorlinks,
    linkcolor={red!50!black},
    citecolor={blue!50!black},
    urlcolor={blue!80!black}
}

\title{\large \bfseries Pulse shape discrimination performance of Inverted Coaxial Ge detectors}

\date{}

\author[1]{\normalsize A. Domula}
\author[2]{\normalsize M. Hult}
\author[3]{\normalsize Y. Kerma\"idic\thanks{Corresponding author, yoann.kermaidic@mpi-hd.mpg.de}}
\author[2]{\normalsize G. Marissens}
\author[3]{\normalsize B. Schwingenheuer}
\author[1]{\normalsize T. Wester}
\author[1]{\normalsize K. Zuber}

\affil[1]{\footnotesize Institut f\"ur Kern- und Teilchenphysik, Technische Universit\"at Dresden, Dresden, Germany}
\affil[2]{\footnotesize Institute for Reference Materials and Measurements, Geel, Belgium}
\affil[3]{\footnotesize Max-Planck-Institut f\"ur Kernphysik, Heidelberg, Germany}
\affil[ ]{\emph{(Dated: November 6, 2017)}}

\begin{document}

 \newcommand{\Ba}{$^{133}$Ba } 
 \newcommand{\Am}{$^{241}$Am } 
  \newcommand{\Amc}{$^{241}$Am, } 
 \newcommand{\Th}{$^{228}$Th } 
 \newcommand{\Thc}{$^{228}$Th, } 
 \newcommand{\Bi}{$^{212}$Bi } 
 \newcommand{\Tl}{$^{208}$Tl } 
 \newcommand{\Ge}{$^{76}$Ge } 

\twocolumn[
\begin{@twocolumnfalse}
	\maketitle
		\thispagestyle{empty}
	\vspace{-2.5cm}
	\begin{abstract}
		We report on the characterization of two inverted coaxial Ge detectors in the context of being employed in future \Ge neutrinoless double beta ($0\nu\beta\beta$) decay experiments.  
		It is an advantage that such detectors can be produced with bigger Ge mass as compared to the planar Broad Energy Ge detectors (BEGe) that are currently used in the GERDA $0\nu\beta\beta$ decay experiment. This will result in lower background for the search of $0\nu\beta\beta$ decay due to a reduction of cables, electronics and holders. 
		The measured resolution near the \Ge Q-value at 2039 keV is 2.5 keV and their pulse-shape characteristics are similar to BEGe-detectors. 
		It is concluded that this type of Ge-detector is suitable for usage in \Ge $0\nu\beta\beta$ decay experiments.
	\end{abstract}
	\vspace{0.5cm}
\end{@twocolumnfalse}
]
\saythanks

\section{Introduction}

Germanium detectors are best suited to measure precisely the energy of MeV scale $\gamma$ rays.
In the past, large high purity Ge detectors with high detection efficiency had a semi-coaxial geometry and a mass of several kg \cite{HD-Moscow,IGEX}.
Point contact detectors, like Broad Energy Germanium (BEGe) detector \cite{BEGe2015} have a lower mass ($\sim$ 0.7 kg) for operational voltage below 5 kV but exhibit a smaller capacitance hence better energy resolution.
In addition, the analysis of the time profile of the BEGe detector signals, called pulse shape analysis \cite{GERDA2013}, allows a powerful discrimination between single or multiple energy depositions inside the crystal or from surface events.
This feature is used in the search for neutrinoless double beta ($0\nu\beta\beta$) decay of \Ge to reject background events with a high efficiency.

The lower limit of the \Ge $0\nu\beta\beta$ ($Q_{\beta\beta} = 2039.061 \pm 0.007 ~\textrm{keV}$ \cite{Qbb2010}) half life has recently been established by the \textsc{GERDA} collaboration at $T_{1/2}^{0\nu}(^{76}\textrm{Ge}) > 5.3 \times 10^{25}$ years (90\% C.L.) \cite{Nature2017}.
In order to further improve this limit, reduction of the background can be achieved by lowering the radioactive mass surrounding the detector (cables, holders).
The recently designed inverted coaxial detectors \cite{Cooper2011} combine the advantages of point contact detectors with larger mass by featuring a well on the opposite side of the contact. 
In this paper the pulse shape analysis of two such commercially available detectors from Canberra, called Small Anode Germanium (SAGe) well detector \cite{Canberra}, are characterized with the aim to study their compatibility with $0\nu\beta\beta$ decay experiment requirements.

\section{Experimental setup}

\begin{figure}
	\centering
	\includegraphics[scale=0.42]{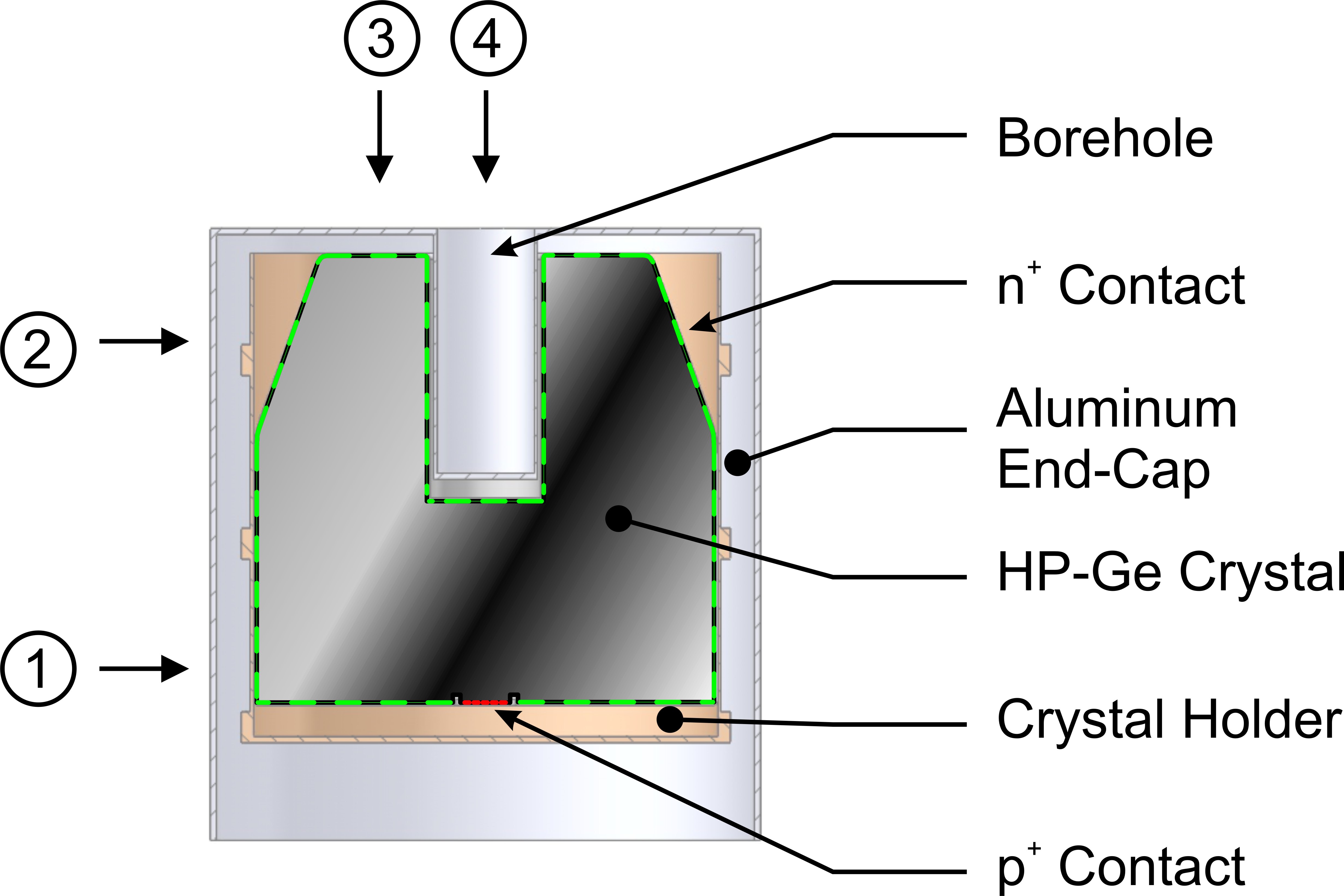}
	\caption{Sketch of the detector "Det-X" environment including the crystal holder and the aluminum end-cap.
		The numbers refer to various source positions used to characterize the HPGe detectors.}
	\label{fig:leadcastle}
\end{figure}

\begin{figure}
	\centering
	\includegraphics[scale=0.5]{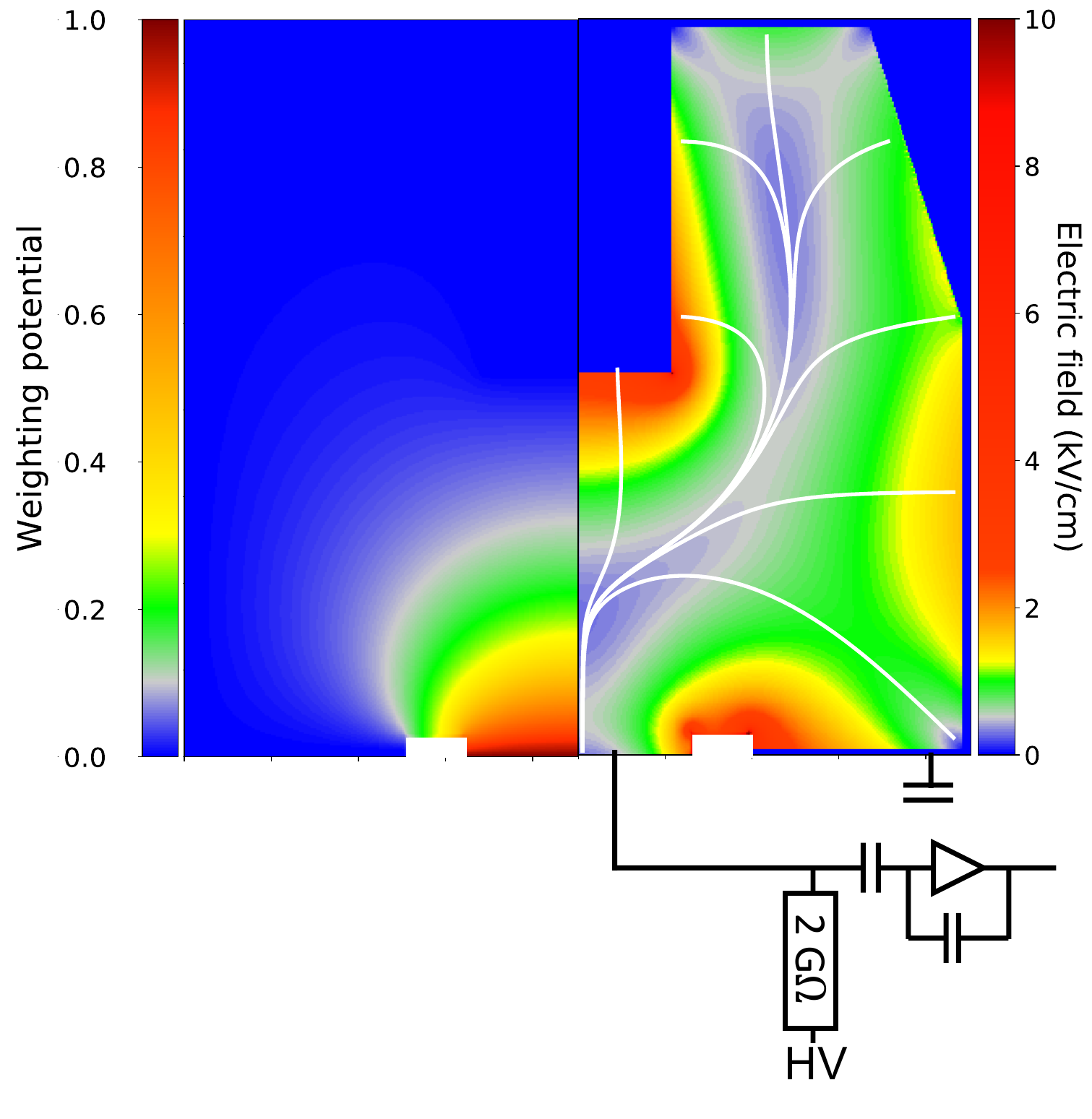}
	\caption{Weighting potential (left) and simulated total electrical field distribution (right) of Ge-14 detector for an impurity concentration of -1.0 to -0.7 $\times 10^{10} \ \textrm{cm}^{-3} $  from the bottom to the top.
		The white lines show some simulated charge carrier paths using ADL software \cite{Bruyneel2016}.}
	\label{fig:weighting_potential}
\end{figure}

The characterization measurements of two inverted coax detectors took place in the HADES underground laboratory in Mol, Belgium (ref. GSW-425P \cite{HADES}) and in the Felsenkeller underground laboratory \cite{DresdenLab} in Dresden (ref. GSW-200P).
With an active mass of 2.6 kg and a volume of 425 cc, the latter detector, referred to as "Ge-14" detector in the following, is the heaviest detector currently available from the manufacturer and works at a reasonable operational voltage while "Det-X" has an active mass of 1.4 kg and a volume of 200 cc.
Each crystal was installed inside a vacuum cryostat located in the center of a lead castle with an inner copper shell (see the HADES setup on Fig. \ref{fig:leadcastle}) for shielding from external radiation.
Similarly to BEGe detectors deployed in \textsc{GERDA}, these crystals are made of p-type high purity Ge.
A high negative potential of -4000 V is applied to the small p+ contact while the rest of the outer surface, covered with a Li-drifted n+ contact, is grounded.
The electric field, which results from the contribution of both the applied high voltage and the intrinsic space charge distribution, is shown in Fig. \ref{fig:weighting_potential} together with the weighting potential \cite{Shockley1938}.
As a consequence of the E-field profile, holes always drift along the same path near the p+ contact ("funnel effect").
Also, the weighting potential features a very low gradient everywhere in the detector except in the vicinity of the p+ contact.
The p+ electrode is AC coupled to a charge sensitive amplifier, delivered by Canberra, for signal readout.
The signal is then digitized with a Flash ADC at a sampling frequency of 100 MHz.
For each triggered signal we record 4000 samples at full FADC resolution, also we save 4000 samples at 25 MHz in order to measure the signal baseline and time constant before and after the trigger, respectively.\\

Two sources, \Amc and \Thc were used for the characterization.
The first source allow localized energy deposition at the detector surface for checking the detector isotropy.
The \Am source was encapsulated in a copper collimator and then placed at different positions around the detectors.
In the following, we show measurements at position 2 for four angles (0\degree, 90\degree, 180\degree  ~ and 270\degree) and position 4 (cf Fig. \ref{fig:leadcastle}).
The latter source ($^{228}$Th), located at position 4, 10 cm above the crystal, produces a broad energy spectrum up to 2.6 MeV with specific event topologies, used for studying pulse shape discrimination.

\section{Detector characterization}

In the following, we report on the energy resolution of both inverted coaxial detectors and compare it to BEGe detectors.
We require to achieve similar values for integrating these new detectors in a $0\nu\beta\beta$ experiment.

\subsection{Energy resolution}

\begin{figure}
\centering
\includegraphics[scale=0.43]{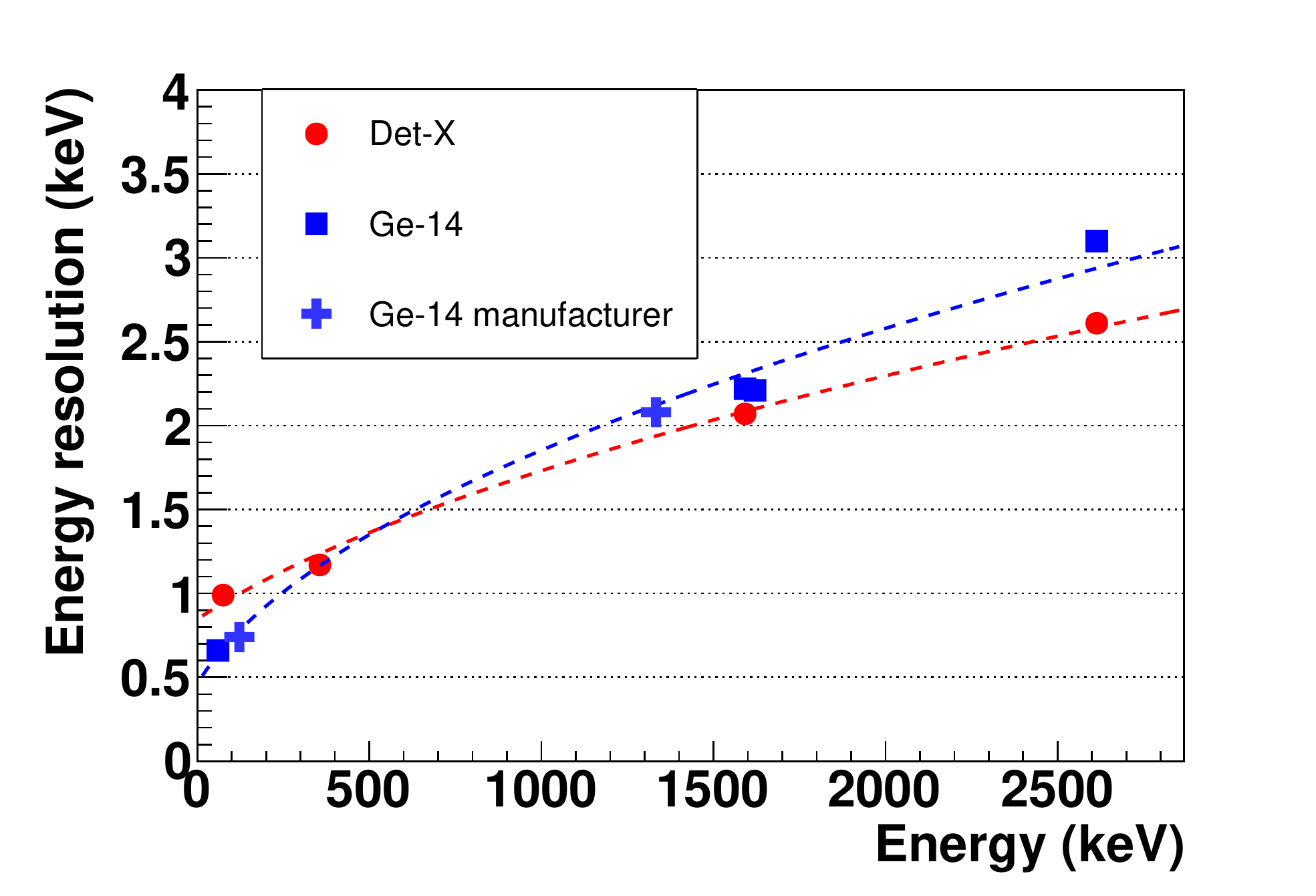}
\caption{Energy resolution as a function of energy for both detectors.
	Manufacturer's data are taken with a $^{57}$Co (122 keV) and a $^{60}$Co (1333 keV line) source.}
\label{fig:EnergyResolution}
\end{figure}

A gaussian fit to the 60 keV \Am line has been performed on top of the Compton background as for the \Tl double escape peak (DEP) at 1592 keV, \Bi full energy peak (FEP) at 1621 keV and \Tl FEP at 2615 keV lines.
We find an energy FWHM resolution of 1.1\% and 1.5\% at 60 keV for Ge-14 and Det-X respectively.
At 1.6 MeV, it reaches 2.22(2.07) keV, i.e. 0.14\%(0.13\%) for the Ge-14(Det-X) detector which is in agreement with the value of 0.16\% at 1.33 MeV reported by Canberra.
The energy dependence of the energy FWHM is plotted for both detectors in Fig. \ref{fig:EnergyResolution}.
From these measurements, we do confer a 20\% energy resolution worsening at $Q_{\beta\beta}$ for the 2.7 kg detector as compared to 0.7 kg BEGe values reported in \cite{BEGe2015}.

\subsection{Signal rise time}

The rise time is an interesting observable for probing the weighting and electric potential homogeneity of a Ge detector close to the p-contact with combining \Am source measurements at different positions.
This parameter is defined as the time interval needed for the charge signal to reach from 5\% to 95\% of its maximum amplitude.

\begin{figure}
	\includegraphics[trim={1.5cm 0 0 0},clip,scale=0.48]{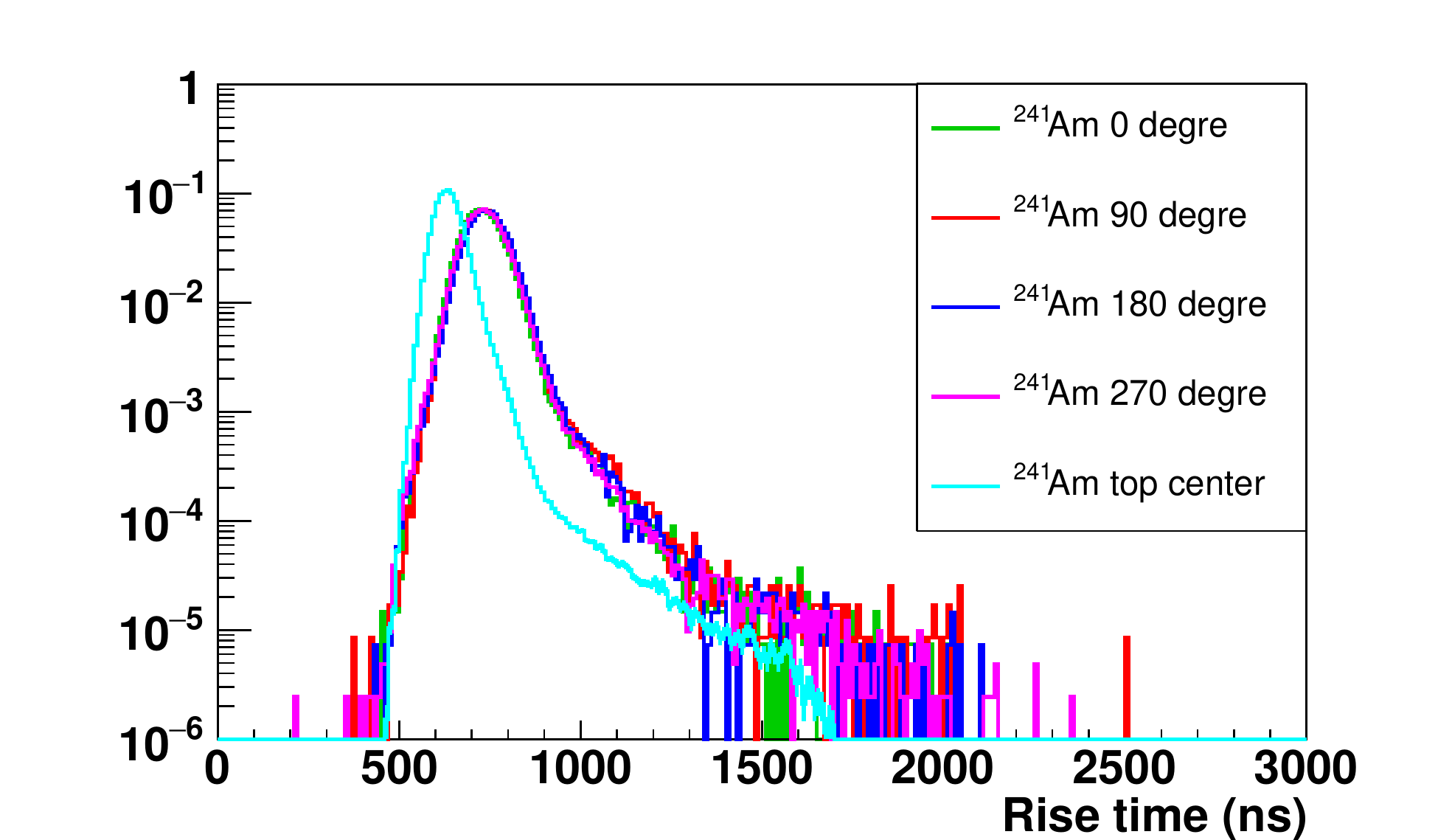}
	\caption{Signal rise time distributions as a function of the \Am source position for the Ge-14 detector.}
	\label{fig:RiseTimeComparison}
\end{figure}

In Fig. \ref{fig:RiseTimeComparison}, we show a comparison of \Am rise time distributions and typical pulses are displayed in Fig. \ref{fig:AveragedPulseComparison}.
No angular dependence is found for the Ge-14 detector.
For pulses located on the side of the 2.7 kg detector, we find an averaged rise time of 750 ns.
Also, due to the non-spherical shape of the weighting potential (cf. Fig. \ref{fig:weighting_potential}), events arising from the well of the detector, (\Am top center), feature an expected lower rise time of about 630 ns.
For comparison, Det-X traces have a much lower averaged rise time of about 420 ns due to its smaller dimensions .
It must be noticed that an angular dependence is found for this detector, leading to a significant 20 ns rise time difference in between $0\degree$ and $180\degree$.
Given the long time charge carriers need to reach the electrode, one expects charge cloud diffusion to worsen the pulse shape analysis performance as compared to BEGe detectors which have typical rise time of the order of 300 ns \cite{Agostini2010}.

\section{Pulse shape discrimination efficiency}
\label{PSD}

\begin{figure}[t!]
	\subfloat[]{\includegraphics[scale=0.35]{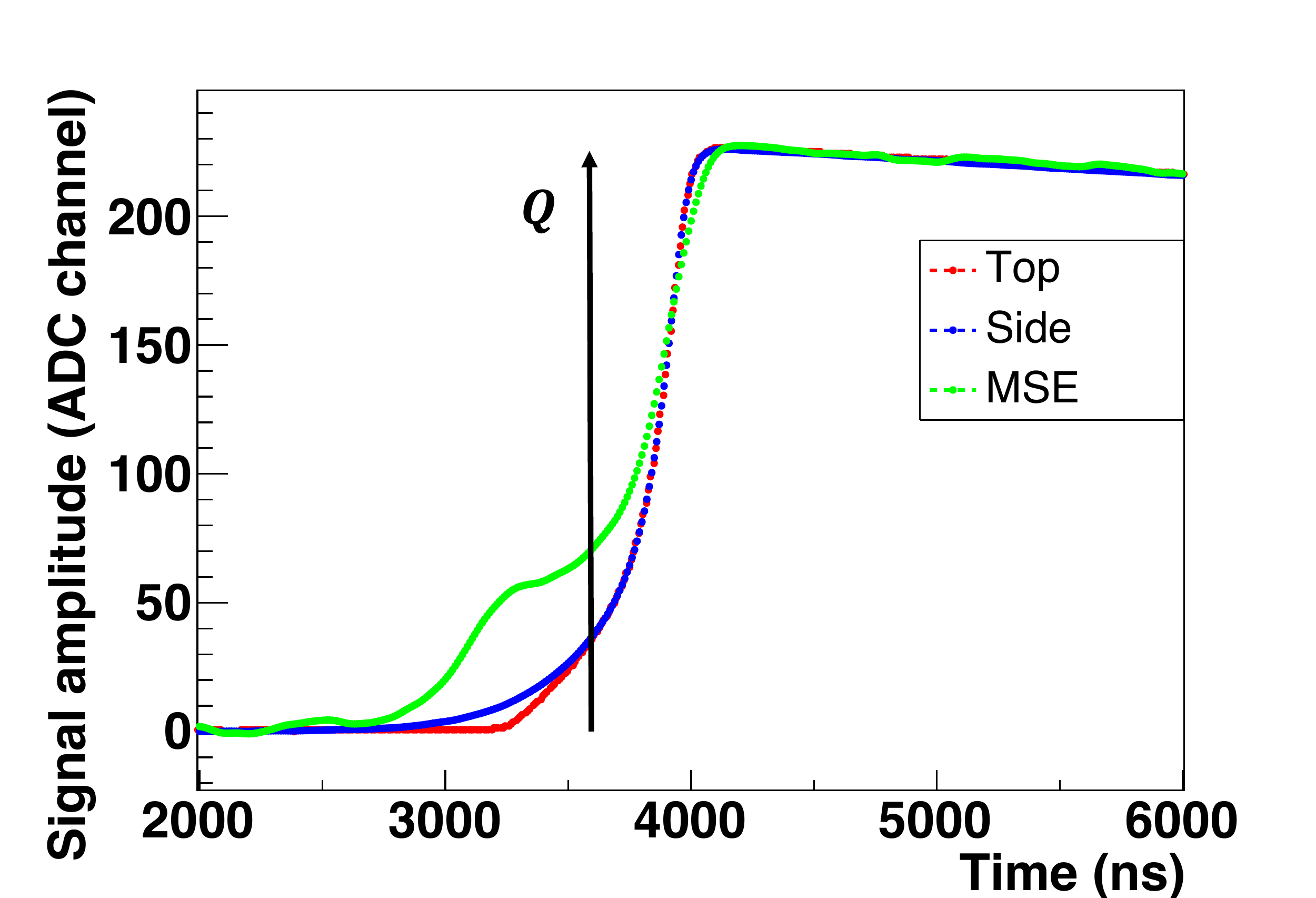}}\\
	\subfloat[]{\includegraphics[scale=0.35]{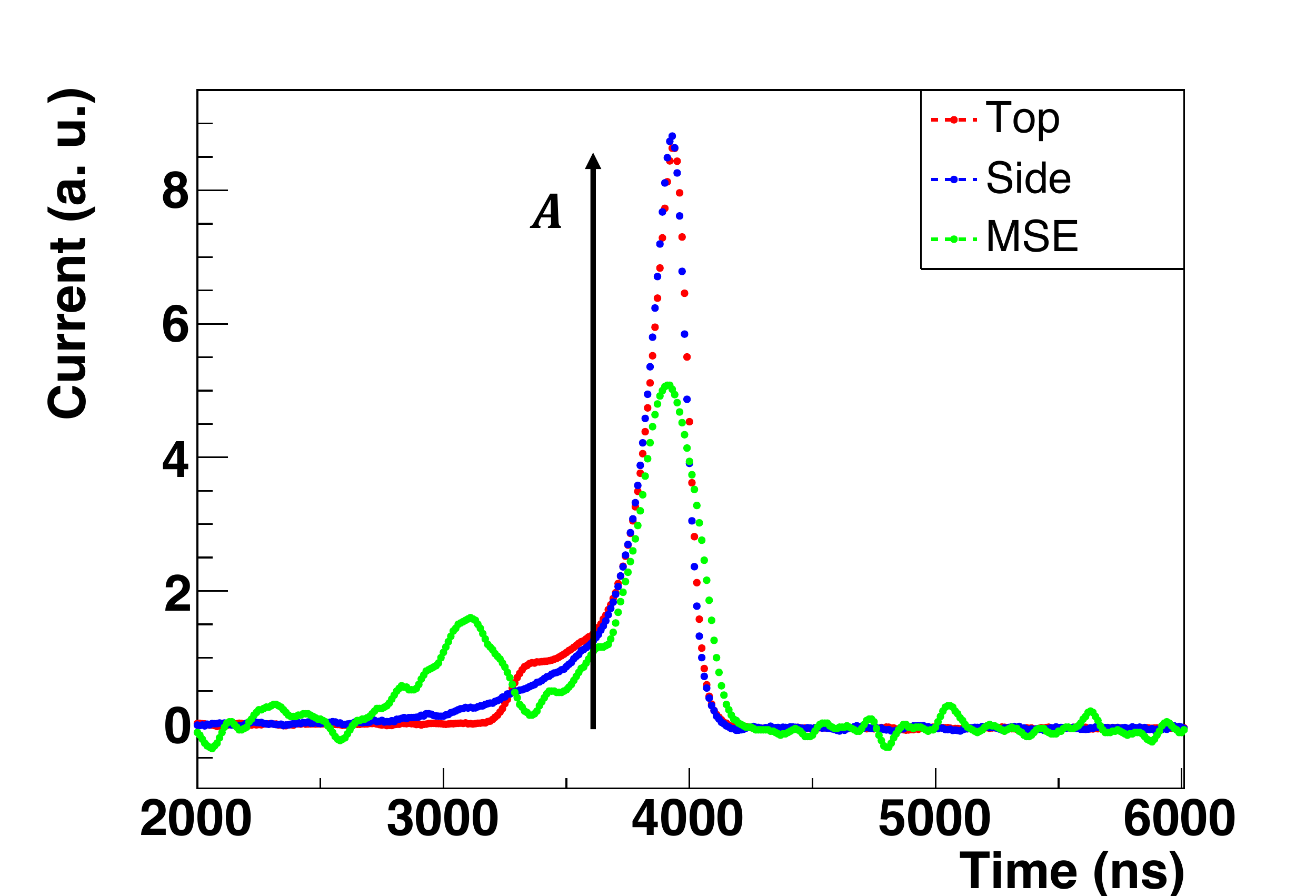}}
	\caption{Example of traces recorded during a measurement using an \Am source.
		On top, comparison for the Ge-14 detector between averaged \Am 59 keV line event pulses for the top (red) and side (blue) (comprising about 3000 pulses) source positions.
		In green, a MSE candidate with the same energy is shown for comparison.
		Below, the same for the corresponding averaged current whose maximal amplitudes depend upon the type of energy deposition (SSE or MSE).}
	\label{fig:AveragedPulseComparison}
\end{figure}

The pulse shape analysis for point-contact Ge detectors in $0\nu\beta\beta$ decay search experiment relies on the ratio between the maximal current amplitude $A$ and the deposited energy $E$.
By virtue of the Shockley-Ramo theorem \cite{He2001}, the current time dependence reads:
\begin{equation}
	I(t) = q \cdot \dot{W}(\vec{r}(t))
\end{equation}
where $q$ is the drifting charge, $\vec{r}(t)$ stands for the charge carriers position at time $t$.
$A$ is therefore proportional to the gradient of the so-called dimensionless weighting potential $W$ of the detector and $E$ to the total collected charge $Q$. 
In Fig. \ref{fig:AveragedPulseComparison}, we compare averaged pulse and corresponding current distributions for Ge-14 from which we estimate $E$ and $A$ respectively for the top and side measurements.

According to Fig. \ref{fig:weighting_potential}, since holes always follow the same path close to the p+ contact (except from a small region close to the electrode) where the weighting potential gradient, i.e. the current, is maximal, one expects from point contact detectors to observe, at a given energy, the same $A$ independent on the position of single energy depositions.
As emphasized in Fig. \ref{fig:AveragedPulseComparison}, the $A/E$ ratio allows to discriminate between single site events (SSE) which are typical $0\nu\beta\beta$ signatures and multi site events (MSE) coming from multiple Compton scattered photons.
Such MSE feature a lower current amplitude as compared to SSE with identical energy.
Only surface events on the p+ contact, like $\alpha$ events, may mimic the SSE signature because here $e^{-}$ drift in a high weighting potential region hence contributing significantly to the charge signal.
However, such events mainly occur at energies above 2 MeV and are therefore not studied in this work.
In the following, we will use the \Tl DEP as a proxy for SSE candidates since in this particular case, pair productions of the 2.6 MeV \Tl line can occur everywhere in the crystal and will deposit energy within a $\textrm{mm}^3$ volume while two 511 keV gamma ray escape the detector.
On the contrary, \Tl SEP features mainly MSE close to the \Ge $Q_{\beta\beta}$ since one of the emitted gammas can make multi-Compton scattering in the detector bulk.
It is worth noting that close to the p+ contact the potential $W$ has a similar shape as for BEGe detectors, we therefore expect fully depleted inverted coaxial detectors to reach the same pulse shape discrimination (PSD) efficiency, i.e. the same MSE rejection power.

\begin{figure}
	\centering
	\includegraphics[scale=0.45]{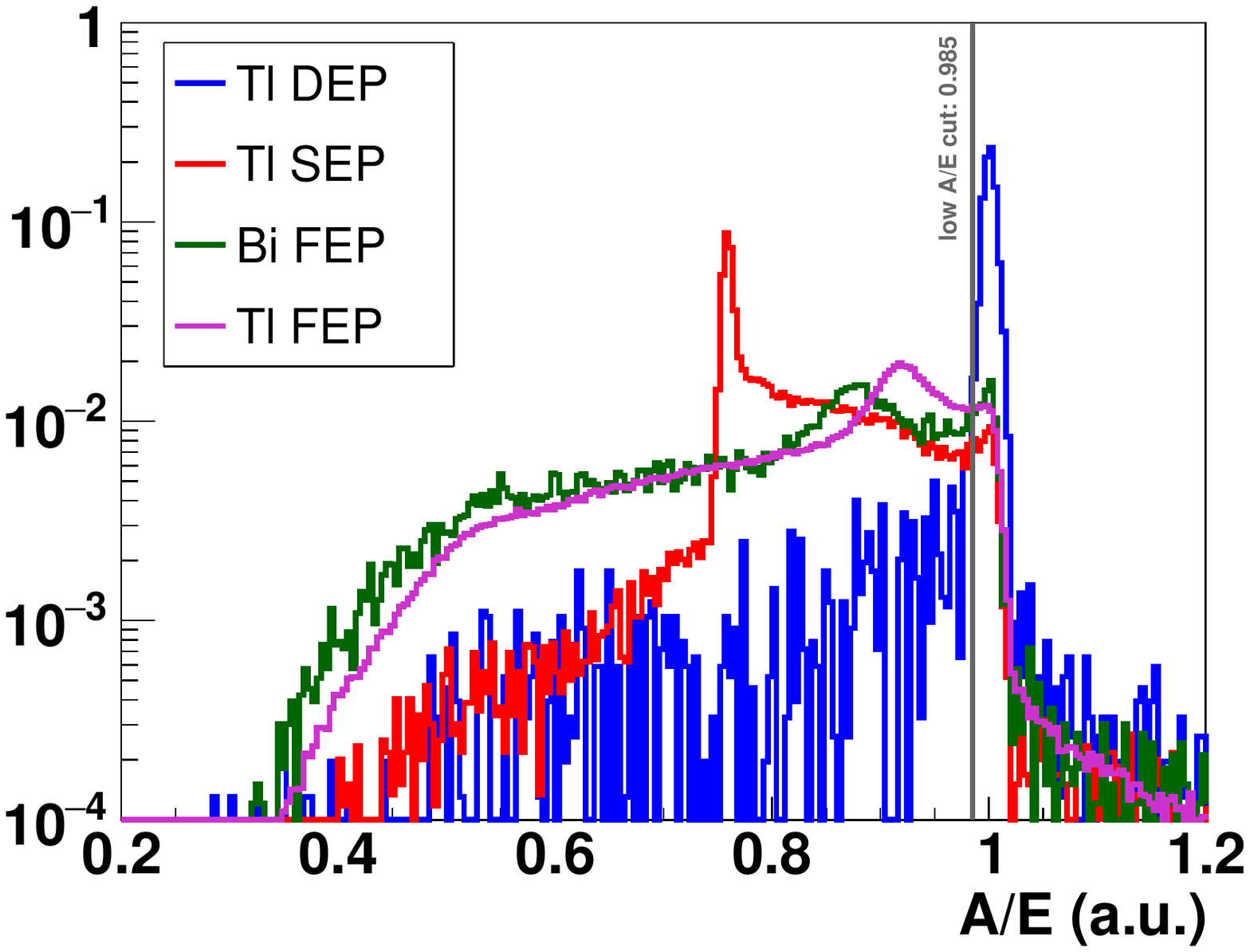}
	\caption{Comparison of the Ge-14 normalized $A/E$ distributions for the \Th source peaks of interest.}
	\label{fig:AoE_peaks}
\end{figure}
\begin{figure}[t!]
	\centering
	\subfloat[]{\includegraphics[scale=0.42]{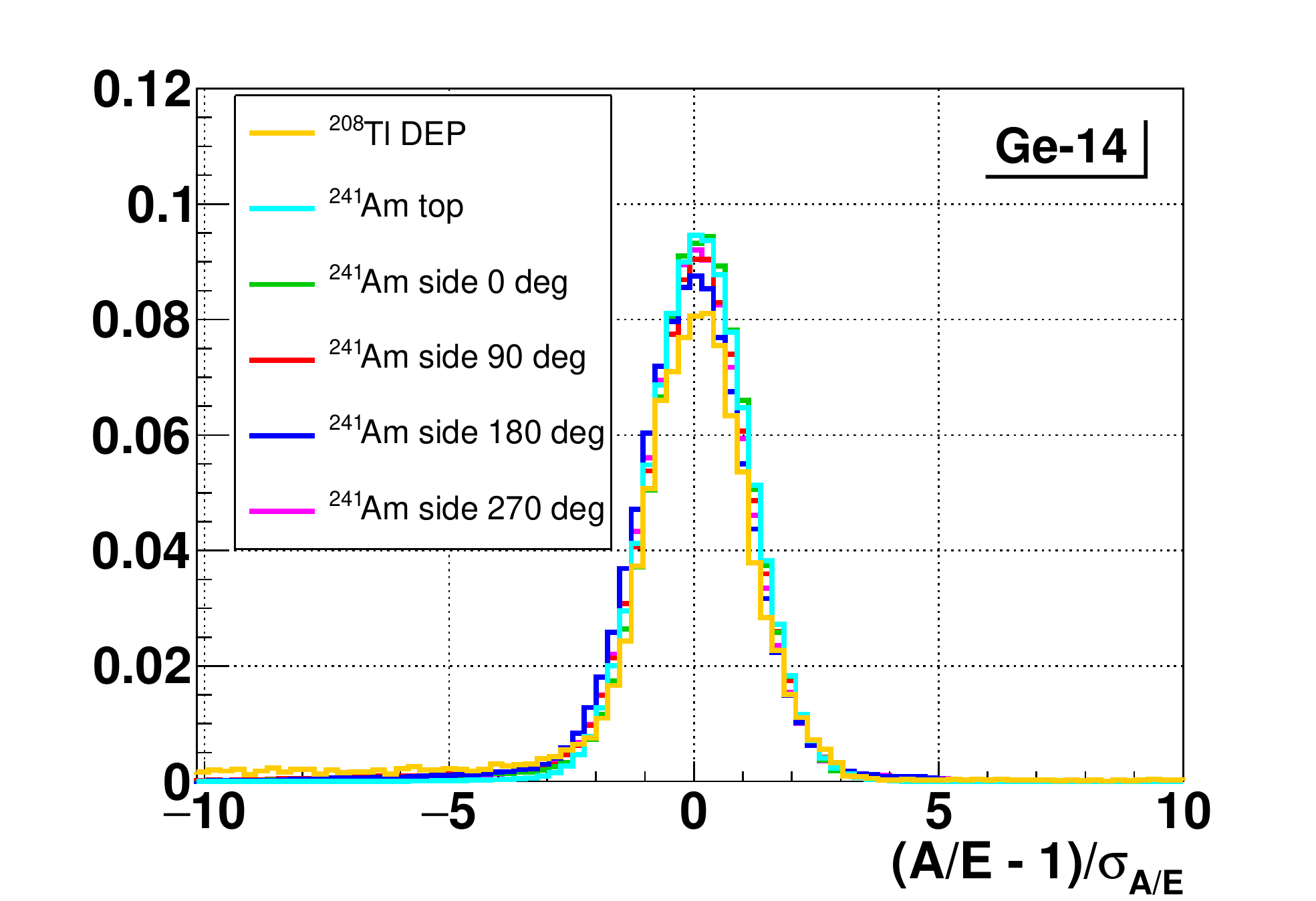}}\\
	\subfloat[]{\includegraphics[scale=0.42]{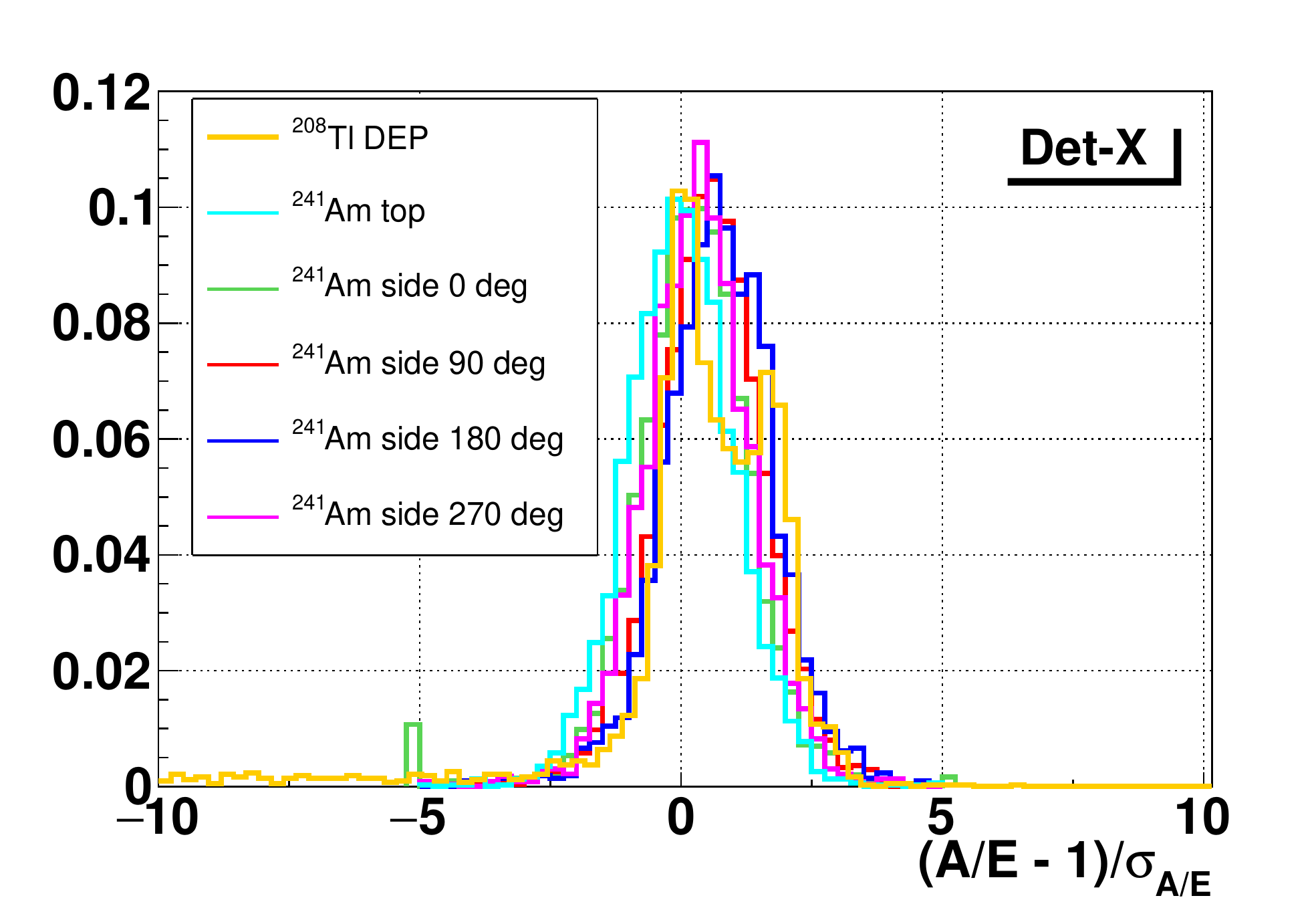}}
	\caption{On top(bottom), comparison of $A/E$ distributions as function of the \Am source position around Ge-14(Det-X) detector.}
	\label{fig:AoE_am}
\end{figure}

\subsection{A/E normalization}

We used the GELATIO software \cite{GELATIO} to obtain $A$ and to reconstruct the energy off-line after applying a moving window average smoothing on each trace.
As already reported in \cite{GERDA2013}, the $A/E$ parameter has a linear energy dependence in our data.
According to our studies with simulations, the energy dependence is mainly caused by diffusion of the charge cloud during the drift.
To correct for this correlation, we normalize it to 1 as follow:
\begin{equation}
A/E = \frac{(A/E)^{\textrm{raw}}}{a \cdot E + b}
\end{equation}
where the parameters $a = 0.15161(2)$ and $b = -1.15(1) 10^{-6}$ for Ge-14 are determined by a linear fit of $A/E$ versus $E$ from three Compton background control regions at 1200 keV, 1700 keV and 2150 keV.
We compare the normalized $A/E$ distributions after Compton background subtraction in Fig. \ref{fig:AoE_peaks} for different event classes of $^{228}$Th decays for the Ge-14 detector.
The subtraction is performed for each energy peak of interest by estimating the number of events in two surrounding energy windows defined as [$-9 \sigma$;$-4.5 \sigma$] and [$4.5 \sigma$;$9 \sigma$] \cite{Wagner2017}.

From this measurement, we find an $A/E$ FWHM for the \Tl DEP of 1.4\% which is dominated by the uncertainty of $A$, coming from the noise of the traces.
This corresponds to an increase of about 40\% as compared to the averaged value of GERDA BEGe detectors \cite{BEGe2015} for the same digital filtering.
Det-X shows an $A/E$ FWHM of 2.3\% for the \Tl DEP.
The factor 1.6 difference between both detectors is explained below.

\subsection{Homogeneity of $A/E$}

We compare in Fig. \ref{fig:AoE_am} the $A/E$ distributions for four \Am source positions (0\degree, 90\degree,180\degree ~ and 270\degree) after background subtraction, using common $a$ and $b$ parameters.
The 60 keV \Am line $A/E$ FWHM resolution are 11\% and 7\% for detector Ge-a4 and Det-X respectively.
A normalization to the $A/E$ resolution $\sigma_{A/E}$ allows to superimpose the \Tl DEP distribution.
For Ge-14, no significant bias or distortion is observed within the $A/E$ resolution, highlighting a proper detector behavior.
On the other hand, the Det-X shows a significant angular dependency of $A/E$ distributions, causing the \Tl DEP distribution to be non-gaussian.
The same observation has been reported in \cite{BEGe2015} for some BEGe detectors operated in vacuum cryostat. This behavior is attributed to positive charged compounds in the groove.
Thus, the inhomogeneity explain the increased $A/E$ width of DEP events.

\begin{figure}
	\centering
	\includegraphics[trim={2.cm 0 0 0},clip,scale=0.38]{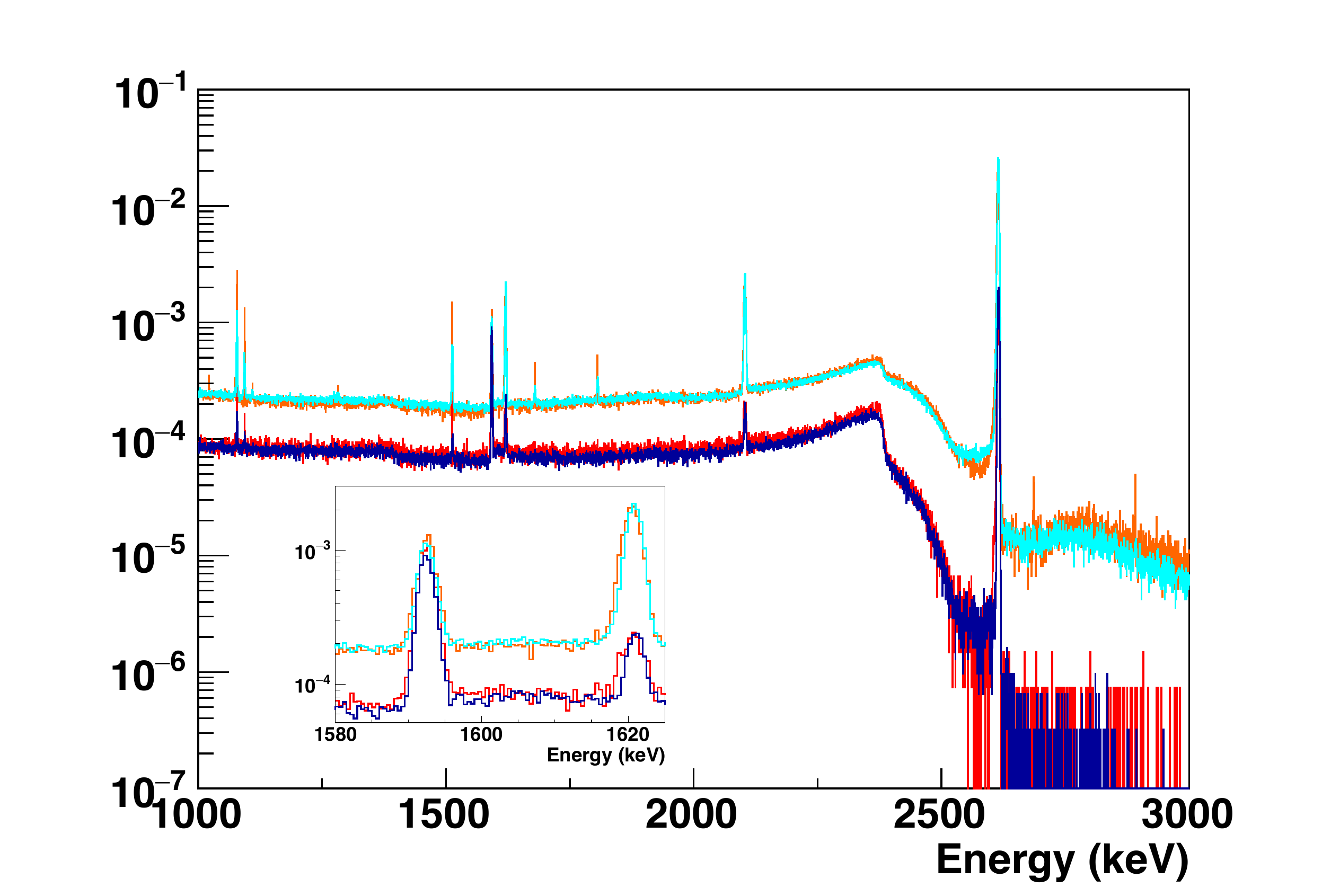}
	\caption{\Th energy spectrum of Ge-14, before(bright blue and orange) and after(dark blue and red) the PSD cut for data and simulation respectively.}
	\label{fig:EnergySpectrumCut}
\end{figure}

\subsection{A/E cuts and survival fraction}

The most important parameter to be determined for $0\nu\beta\beta$ experiments, is the survival fraction of energy peaks featuring mainly SSE (\Tl DEP) or MSE (\Tl SEP) after applying the PSD cut.
A constant PSD cut, displayed in Fig. \ref{fig:AoE_peaks}, has been applied.
The one-sided PSD cut is defined such that 90\% of the \Tl DEP events survive.
Monte-Carlo simulations, based on MaGe \cite{MaGesoftware} and ADL \cite{Bruyneel2016} softwares, have been performed for Ge-14 to compare the expected performances to the data.
The energy spectrum before and after the PSD cut for both data and simulation are shown in Fig. \ref{fig:EnergySpectrumCut}.
It is worth noting that the detector energy resolution has been included in the simulation only to the four energy peaks of interest.
We find a \Tl SEP events survival fraction of 7.0(4)\% in simulation to be compared to 6.2(2)\% in data (a 1.5\%-2\% agreement is found for the \Tl FEP and \Bi FEP respectively).
The survival fractions are summarized in table \ref{tab:survival_fraction}.
In general, they are all compatible with the observed averaged values reported for BEGe detectors in \cite{BEGe2015} for both Ge-14 and Det-X despite the observed anisotropy reported in Fig. \ref{fig:AoE_am} for the latter.
From table \ref{tab:survival_fraction}, it follows that the expected DEP $A/E$ resolution worsening coming from a longer charge carriers drift path for the inverted coaxial detectors is compensated by the geometry change.
This latter difference modifies the discriminating power of the $A/E$ cut as it can be seen by comparing the $A/E$ distributions on Fig. \ref{fig:AoE_peaks} and \cite{BEGe2015}.

\begin{table}[h!]
	\centering
	\begin{tabular}{|c|c|c|c|} 
		\hline
		(\%)       & Ge-14 & Det-X & BEGe  \\ 
		\hline
		\Tl DEP     	& 89.8 	   & 90.2 & 90.0   \\ \hline
		\Bi FEP  		& 9.6 		& 8.4 	&  11.5 \\ \hline		 
$Q_{\beta\beta}$  & 32.7 	 & 35.6 &  37.8 \\ \hline		
		\Tl SEP  		& 6.2 		& 5.5 	& 7.5  \\ \hline
		\Tl FEP  		&  8.6 		& 8.0 	&  7.7  \\ \hline
	\end{tabular}
	\caption{Survival fraction of the one-sided PSD cut of the four peaks of interest and the Compton background suppression in the $Q_{\beta\beta}$ region.
	The uncertainties are all below the percent level.
The BEGe data correspond to the average of reported values in \cite{BEGe2015}}
	\label{tab:survival_fraction}
\end{table}

\section{Conclusion}

We characterized two inverted coaxial detectors specifically in terms of energy resolution and PSD efficiency.
We find performances close to existing BEGe detectors that are successfully deployed in \textsc{GERDA}.
We also performed preliminary benchmark between the data and Monte-Carlo simulations. 
We found an agreement at the percent level between $A/E$ distributions after PSD cut which is of great interest in order to further optimize detector geometry with simulation and to understand the background after cuts in experiments like \textsc{GERDA}.
The main conclusion of this study is that inverted coaxial detectors are suitable for $0\nu\beta\beta$ decay searches and open a door for improving the background index of incoming larger \Ge based experiments.

\paragraph{Acknowledgment}
The transnational access scheme (EUFRAT) of the European Commission's Joint Research Centre in Geel is gratefully acknowledged.
We are also thankful to David Radford for helpful discussions.


\vspace{-1cm}
\renewcommand{\refname}{}



\begin{thebibliography}{10}

\bibitem{HD-Moscow}
H.~Klapdor-Kleingrothaus, et~al.,
\newblock Physics Letters B \textbf{586}, 198 (2004).

\bibitem{IGEX}
C.~E. Aalseth, et~al.,
\newblock Phys. Rev. D \textbf{65}, 092007 (2002).

\bibitem{BEGe2015}
M. Agostini, et~al. (GERDA Collaboration),
\newblock Eur. Phys. J. C \textbf{75}, 39 (2015).

\bibitem{GERDA2013}
M. Agostini, et~al. (GERDA Collaboration),
\newblock Eur. Phys. J. C \textbf{73}, 2583 (2013).

\bibitem{Qbb2010}
B.~J. Mount, M.~Redshaw, and E.~G. Myers.,
\newblock Phys. Rev. C \textbf{81}, 032501 (2010).

\bibitem{Nature2017}
M. Agostini, et~al. (GERDA Collaboration),
\newblock Nature \textbf{544}, 47 (2017).

\bibitem{Cooper2011}
R.~Cooper, et~al.,
\newblock Nucl. Instrum. Methods Phys. Res. Sect. A \textbf{665}, 25 (2011).

\bibitem{Canberra},
C.~\textit{The SAGe Well: A New Revolution in Well and Environmental Counting}.

\bibitem{HADES}
M.~Hult, et~al.,
\newblock Appl. Radiat. Isot., In press (2017).

\bibitem{DresdenLab}
S.~Niese, M.~Koehler, and B.~Gleisberg.,
\newblock Journal of Radioanalytical and Nuclear Chemistry \textbf{233}, 167 (1998).

\bibitem{Shockley1938}
W.~Shockley.,
\newblock Journal of Applied Physics \textbf{9}, 635 (1938).

\bibitem{Bruyneel2016}
B.~Bruyneel, B.~Birkenbach, and P.~Reiter.,
\newblock Eur. Phys. J. A \textbf{52}, 70 (2016).

\bibitem{Agostini2010}
M.~Agostini, et~al.,
\newblock J. Instrum. \textbf{6}, P03005 (2011).

\bibitem{He2001}
Z.~He.,
\newblock Nucl. Instrum. Methods Phys. Res. Sect. A \textbf{463}, 250 (2001).

\bibitem{GELATIO}
M.~Agostini, et~al.,
\newblock J. Instrum. \textbf{6}, P08013 (2011).

\bibitem{Wagner2017}
V.~Wagner.,
\newblock PhD thesis, Ruperto-Carola university of Heidelberg (2017).

\bibitem{MaGesoftware}
M.~Bauer, et~al.,
\newblock Journal of Physics: Conference Series \textbf{39}, 362 (2006).

\end{thebibliography}

\end{document}